\documentclass[aps,prc,twocolumn,showpacs,superscriptaddress,footinbib]{revtex4-1}

\usepackage{graphicx}
\usepackage{xcolor}
\usepackage{dcolumn}
\usepackage{bm}
\usepackage{enumerate}
\usepackage{amsmath}

\usepackage[normalem]{ulem}  

\begin{document}


\title{$B(E2;2^{+}_{1}\rightarrow0^{+}_{1})$ anomaly in $^{166}$Os}

\author{Chen-guang Zhang}
\email{391860206@qq.com}
\affiliation{College of Physics, Tonghua Normal University, Tonghua 134000, People's Republic of China}

\author{Suo-chang Jin}
\email{18243725176@163.com}
\affiliation{Department of Physics,College of Physics, Yanbian University, Yanji, Jilin 133002, People's Republic of China}

\author{Tie Wang}
\email{twang@ybu.edu.cn}
\affiliation{Department of Physics,College of Physics, Yanbian University, Yanji, Jilin 133002, People's Republic of China}

\author{Tao Wang}
\email{suiyueqiaoqiao@163.com}
\affiliation{College of Physics, Tonghua Normal University, Tonghua 134000, People's Republic of China}

\date{\today}

\begin{abstract}
\textbf{Abstract:} Recently, the very small $B(E2;2_{1}^{+}\rightarrow0_{1}^{+})$ value was found in $^{166}$Os, which is 7(4) W.u. \textbf{and} much smaller than the values of adjacent nuclei $^{168,170}$Os, 74(13) W.u. and 97(9) W.u.. With the help of a new technique ``SU(3) analysis" and a new explanatory framework of the B(E2) anomaly, it was found that, the $B(E2;2_{1}^{+}\rightarrow0_{1}^{+})$ anomaly really can exist. In this paper, we discuss the $B(E2;2_{1}^{+}\rightarrow0_{1}^{+})$ anomaly in $^{166}$Os for the first time. Four results are used to fit the experimental data in $^{166,168,170}$Os successfully. This implies that the level-crossing or level-anticrossing explanation is more reasonable.

\end{abstract}

\maketitle

\section{Introduction}

If the ratio $E_{4/2}=E_{4_{1}^{+}}/E_{2_{1}^{+}}$ of the energies of the 4$_{1}^{+}$ and 2$_{1} ^{+}$ states is larger than 2.0, it is usually regarded as a signal for the \textbf{emergence} of various collective excitations. Meanwhile, the ratio $B_{4/2}=B(E2;4_{1}^{+}\rightarrow2_{1}^{+})/B(E2;2_{1}^{+}\rightarrow0_{1}^{+})$ of the E2 transitions $B(E2;4_{1}^{+}\rightarrow2_{1}^{+})$ and $B(E2;2_{1}^{+}\rightarrow0_{1}^{+})$ is usually larger than 1.0. \textbf{If $E_{4/2}>2.0$ but $B_{4/2}<1.0$, this anomalous phenomenon is called B(E2) anomaly.} Experimentally, the B(E2) anomaly has been discovered in $^{112,114}$Xe \cite{Xe112,Xe114}, $^{114}$Te \cite{Moller05}, $^{168,170}$Os \cite{Garahn16,Goasduff19}, $^{166}$W \cite{Joss17}, $^{172}$Pt \cite{Cederwall18}, \textbf{and }in the even-odd nuclei $^{167,169}$Os \cite{Cederwall21,Cederwall25} and $^{119}$Te \cite{Cederlof25}. Thus the \textbf{origin} of the B(E2) anomaly \textbf{challenges various nuclear structure theories.}

The interacting boson model (IBM) \textbf{was established} to explain the collective characteristics of nuclei \cite{Iachello75,Iachello87}. Nucleon-pairs with angular momentum $L=0$ and $L=2$ \textbf{are regarded as $s$ and $d$ bosons}. The IBM possesses the U(6) symmetry and has four dynamical symmetry limits: (1) the U(5) symmetry limit is used to describe the surface vibrations of the spherical shape; (2) the SU(3) symmetry limit is used to describe the rotational spectra of the prolate shape; (3) the O(6) symmetry limit is used to describe the $\gamma$-soft rotational mode; \textbf{and} (4) the $\overline{\textrm{SU(3)}}$ symmetry limit is used to describe the rotational spectra of the oblate shape \cite{Jolie01}. Thus the IBM can be also used to describe various shape phase transitions between the different shapes \cite{Warner02,Casten06,Casten07,Bonatsos09,Casten09,Jolie09,Casten10,Jolos21,Fortunato21,Cejnar21,Bonatsos24,Jolie00,Cejnar03,Iachello04,Wang08}.

\textbf{Recently}, an extended version of the IBM by incorporating the SU(3) symmetry higher-order interactions (\textbf{SU3-IBM for short}) was proposed , \textbf{which} combines previous IBM concepts and the SU(3) correspondence of the rigid triaxial rotor \cite{Isacker85,Draayer87,Draayer88,Isacker00,Kota20}. In this new model, the SU(3) symmetry governs all the quadrupole deformations \textbf{including the oblate shape}.

The SU3-IBM \textbf{has been successfully} used to explain the B(E2) anomaly \textbf{with higher-order interactions} \cite{Wang20,Zhang22,Wangtao,Zhang24,Pan24,Zhang25,Zhang252,Teng25,Cheng25,Li25}, \textbf{to resolve} the Cd puzzle \cite{Heyde11,Heyde16,Garrett18} \textbf{with a newly proposed spherical-like $\gamma$-soft spectra} \cite{Wang22,Wang25,WangPd106,Zhao25}, \textbf{to more correctly describe} the asymmetric prolate-oblate shape phase transition in the Hf-Hg region \cite{Fortunato11,Zhang12,Wang23}, \textbf{to} more accurately describe \textbf{the} $\gamma$-soft behaviors in $^{196}$Pt \cite{WangPt}, \textbf{to} explain the unique boson number odd-even phenomena in $^{196-204}$Hg \textbf{which was a prediction of the new model for the oblate shape} \cite{Zhang12,WangHg}, \textbf{to} describe the E(5)-like spectra of $^{82}$Kr \textbf{in a new way} \cite{Zhou23}, \textbf{and} \textbf{to} describe the rigid triaxiality of $^{166}$Er in an excellent way \cite{ZhouEr} \textbf{which was predicted by Otsuka \emph{et al.}} \cite{Otsuka19,Otsuka21,Otsuka}. These findings support the validity of the SU3-IBM.

\textbf{In the level-crossing mechanism \cite{Wang20,Cheng25}, the B(E2) anomaly results from the different collectivity of the two $2_{1}^{+}$ and $4_{1}^{+}$ states for in the SU(3) symmetry limit they have different SU(3) irreducible representation (irrep). In the rigid triaxial mechanism \cite{Zhang22}, this comes from one specific triaxial deformation and finite-$N$ effect. In this SU(3) symmetry limit, only one SU(3) irrep is considered. These two mechanisms describe the B(E2) anomaly from two different viewpoints, and further revealing the relationships between them is important. When discussing a rigid triaxial rotor, the higher-order interactions also generate level-crossing or level-anticrossing between different deformations, see the discussions in \cite{Cheng25}. Recently, a new mechanism was proposed in \cite{Pan24}, and in a recent paper \cite{Li25}, this new mechanism can combine the level-crossing mechanism into a general explanatory framework for the B(E2) anomaly. }

Most surprisingly, the $B(E2;2_{1}^{+}\rightarrow0_{1}^{+})$ value in $^{166}$Os was found to be very small, 7(4) W.u. \cite{Stolze21}, which is much smaller than the values in the neighbouring nuclei $^{168,170}$Os, 74(13) W.u. and 97(9) W.u.. In $^{168,170}$Os, the $B_{4/2}$ values are smaller than 1.0. Thus a good theory \textbf{should} describe not only the B(E2) anomaly in $^{168,170}$Os but also the $B(E2;2_{1}^{+}\rightarrow0_{1}^{+})$ anomaly in $^{166}$Os.

\begin{figure}[tbh]
\includegraphics[scale=0.33]{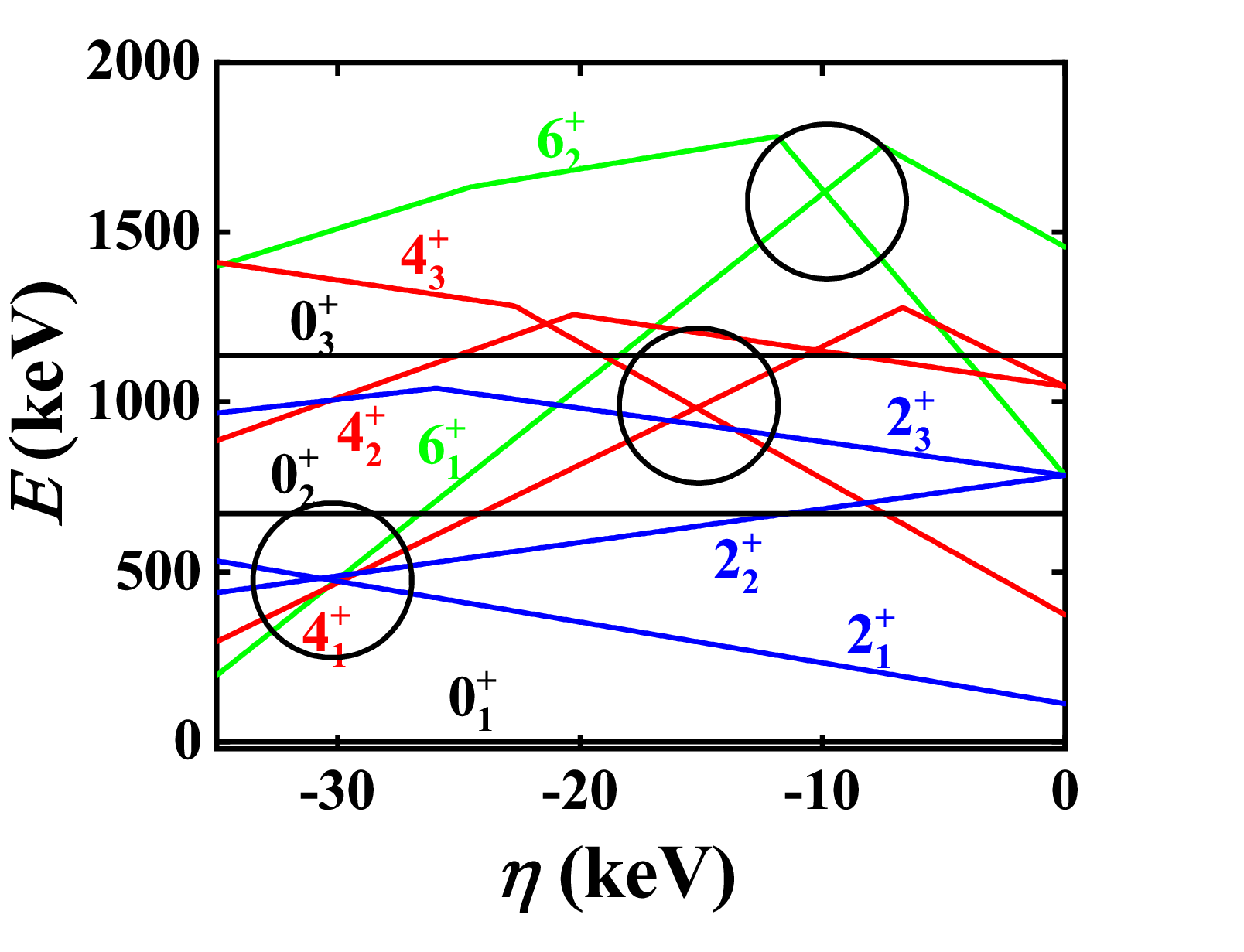}
\includegraphics[scale=0.33]{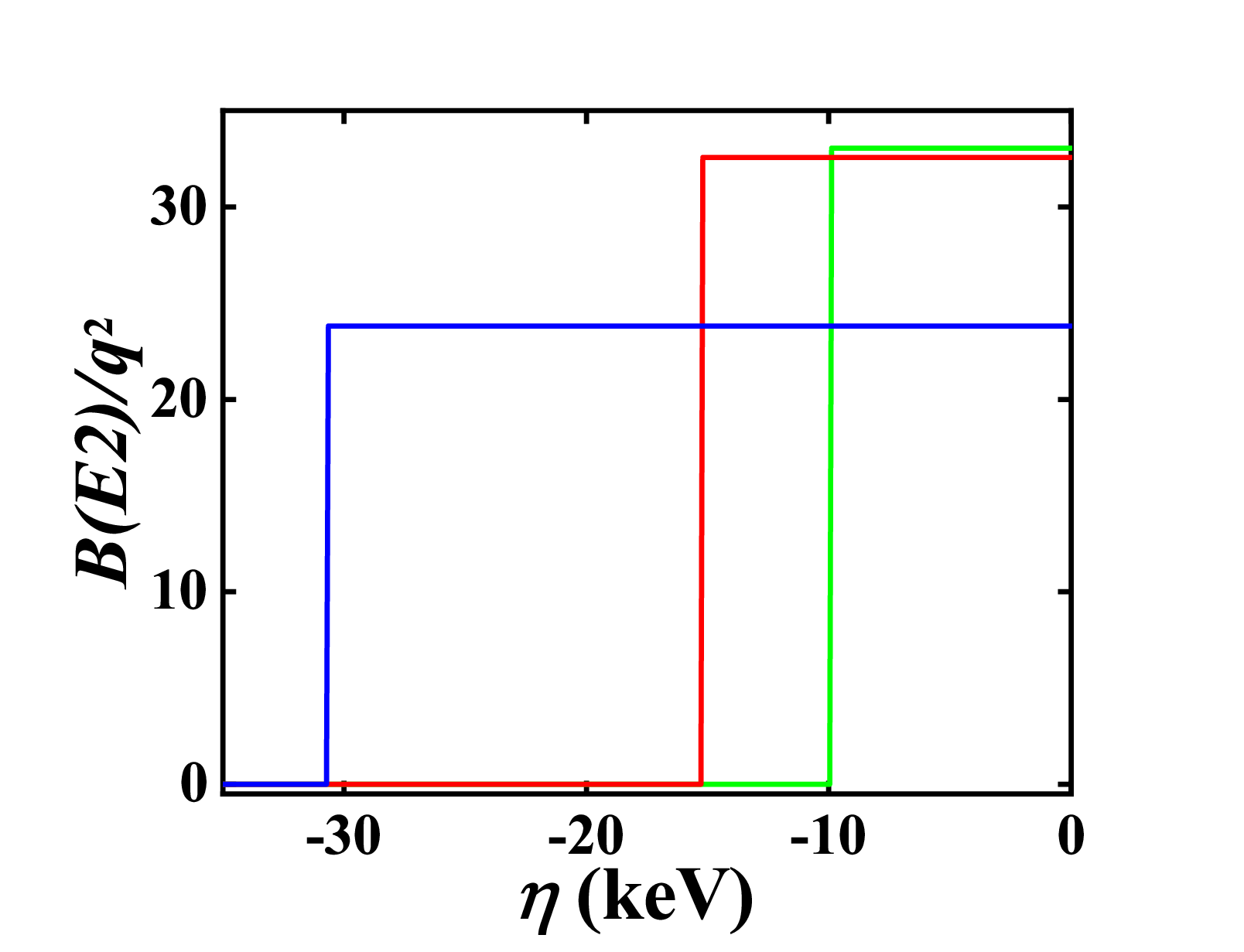}
\caption{(a) The evolutional behaviors of the partial low-lying levels as a function of $\eta$; (b) The evolutional behaviors of the $B(E2; 2_{1}^{+}\rightarrow 0_{1}^{+})$ (blue line) , $B(E2; 4_{1}^{+}\rightarrow 2_{1}^{+})$ (red line), $B(E2; 6_{1}^{+}\rightarrow 4_{1}^{+})$ (green line) as a function of $\eta$. The parameters are deduced from \cite{Wang20}.}
\end{figure}

In this paper, we discuss this problem \textbf{for the first time}. With the help of the SU(3) analysis \cite{Cheng25} and the general explanational framework \cite{Li25}, four results are explored to fit the $B(E2;2_{1}^{+}\rightarrow0_{1}^{+})$ anomaly in $^{166}$Os and the B(E2) anomaly in $^{168,170}$Os. These results obtained fit well. It implies that the level-crossing or level-anticrossing explanation can have greater explanatory power.

\section{A brief introduction to SU(3) analysis, level-crossing and level-anticrossing}

\begin{figure}[tbh]
\includegraphics[scale=0.33]{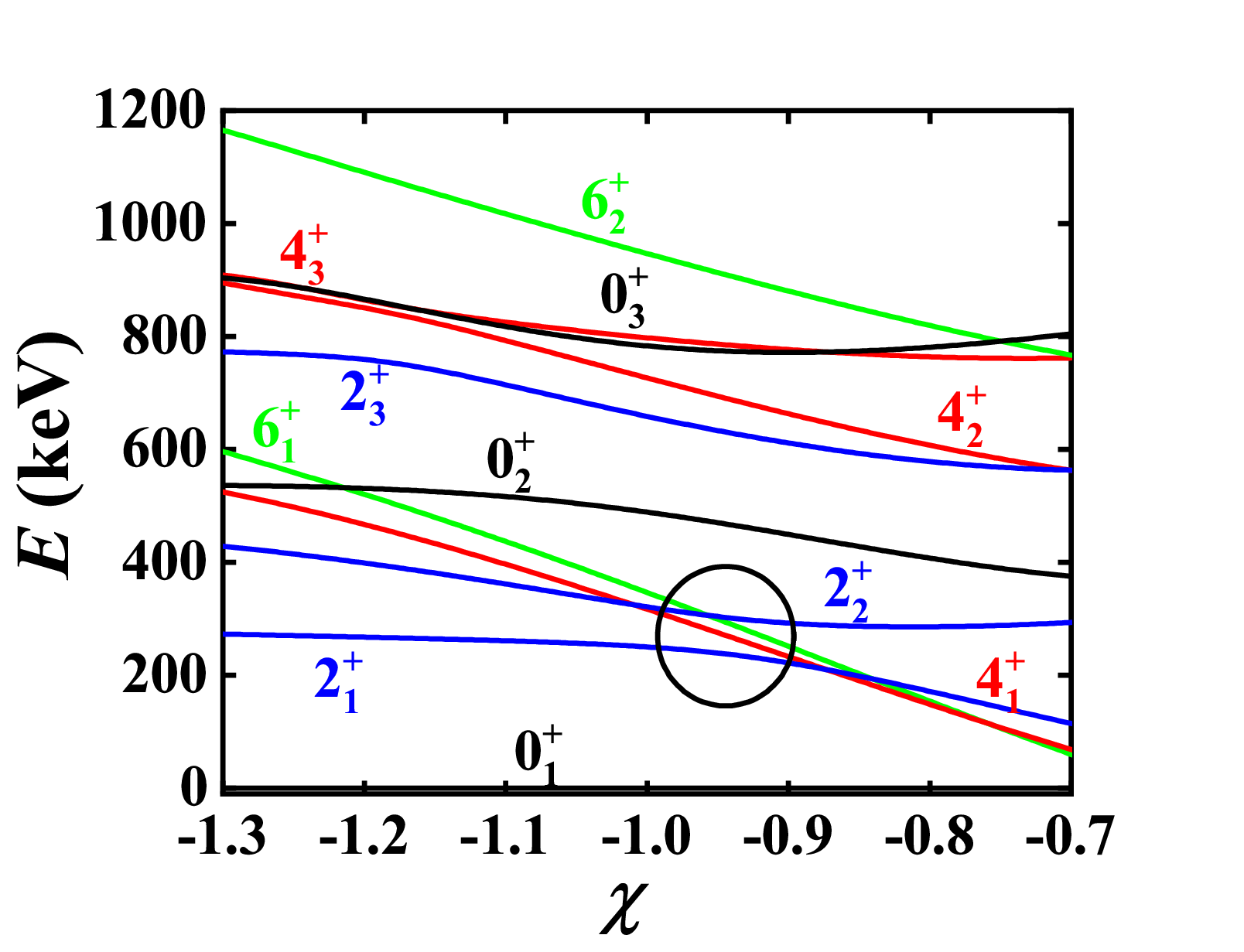}
\includegraphics[scale=0.33]{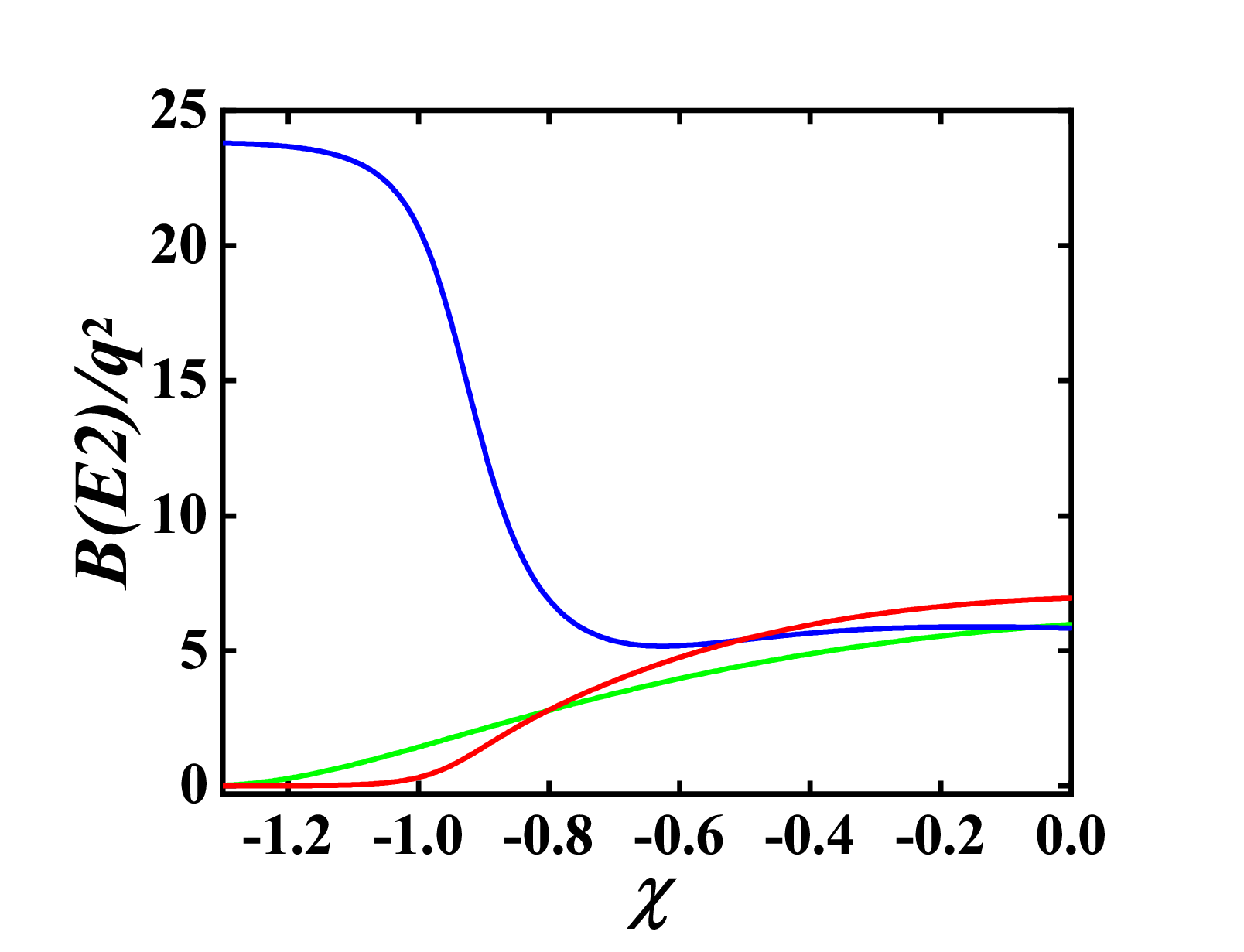}
\caption{(a) The evolutional behaviors of the partial low-lying levels as a function of $\chi$; (b) The evolutional behaviors of the $B(E2; 2_{1}^{+}\rightarrow 0_{1}^{+})$ (blue line) , $B(E2; 4_{1}^{+}\rightarrow 2_{1}^{+})$ (red line), $B(E2; 6_{1}^{+}\rightarrow 4_{1}^{+})$ (green line) as a function of $\chi$. The parameters are deduced from \cite{Wang20}.}
\end{figure}

\textbf{Since the new model has not been proposed for a long time and the related concepts of SU(3) analyis are newly given \cite{Cheng25}, a brief introduction to SU(3) ananlyis, level-crossing and level-anticrossing is necessary.}

\begin{figure}[tbh]
\includegraphics[scale=0.33]{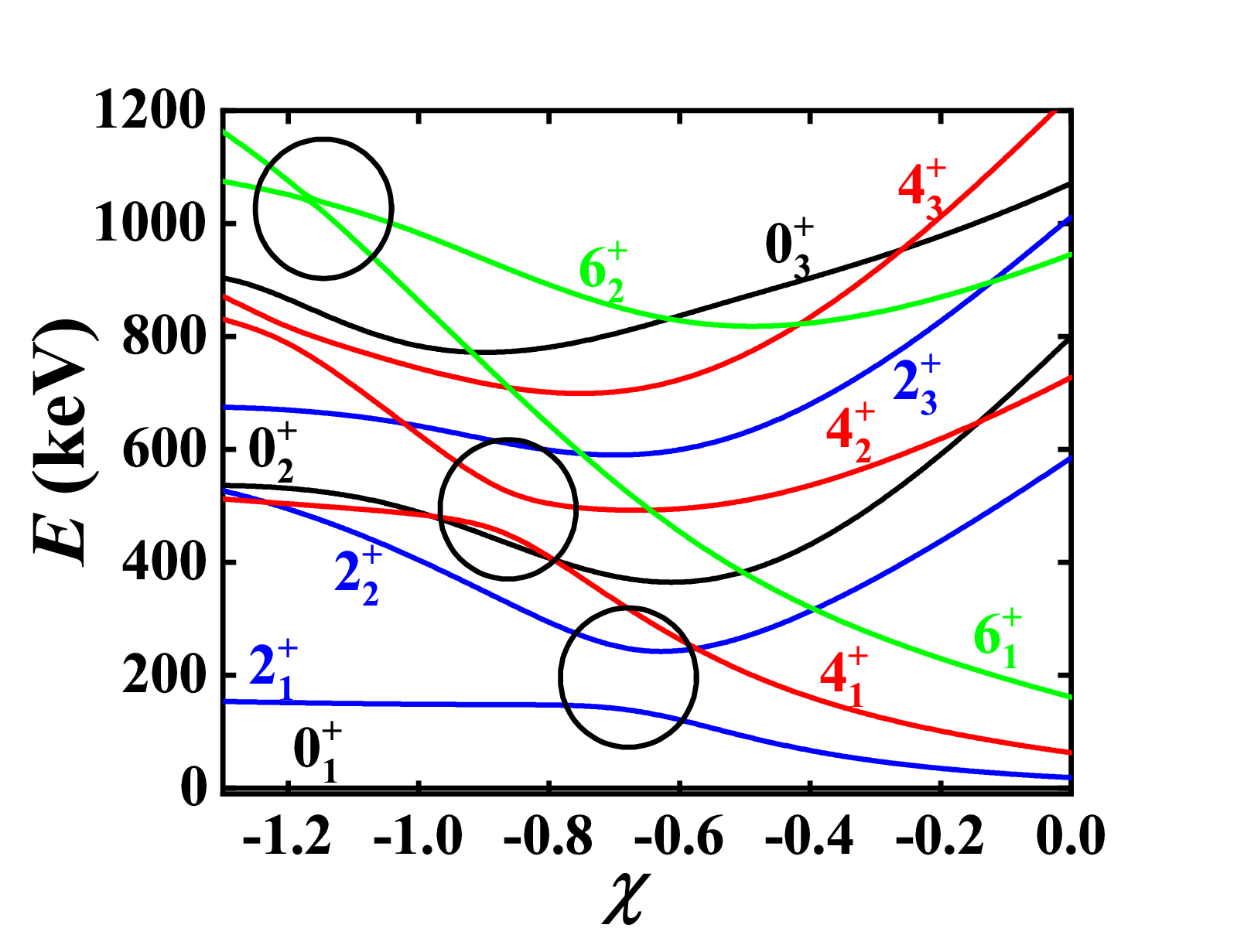}
\includegraphics[scale=0.33]{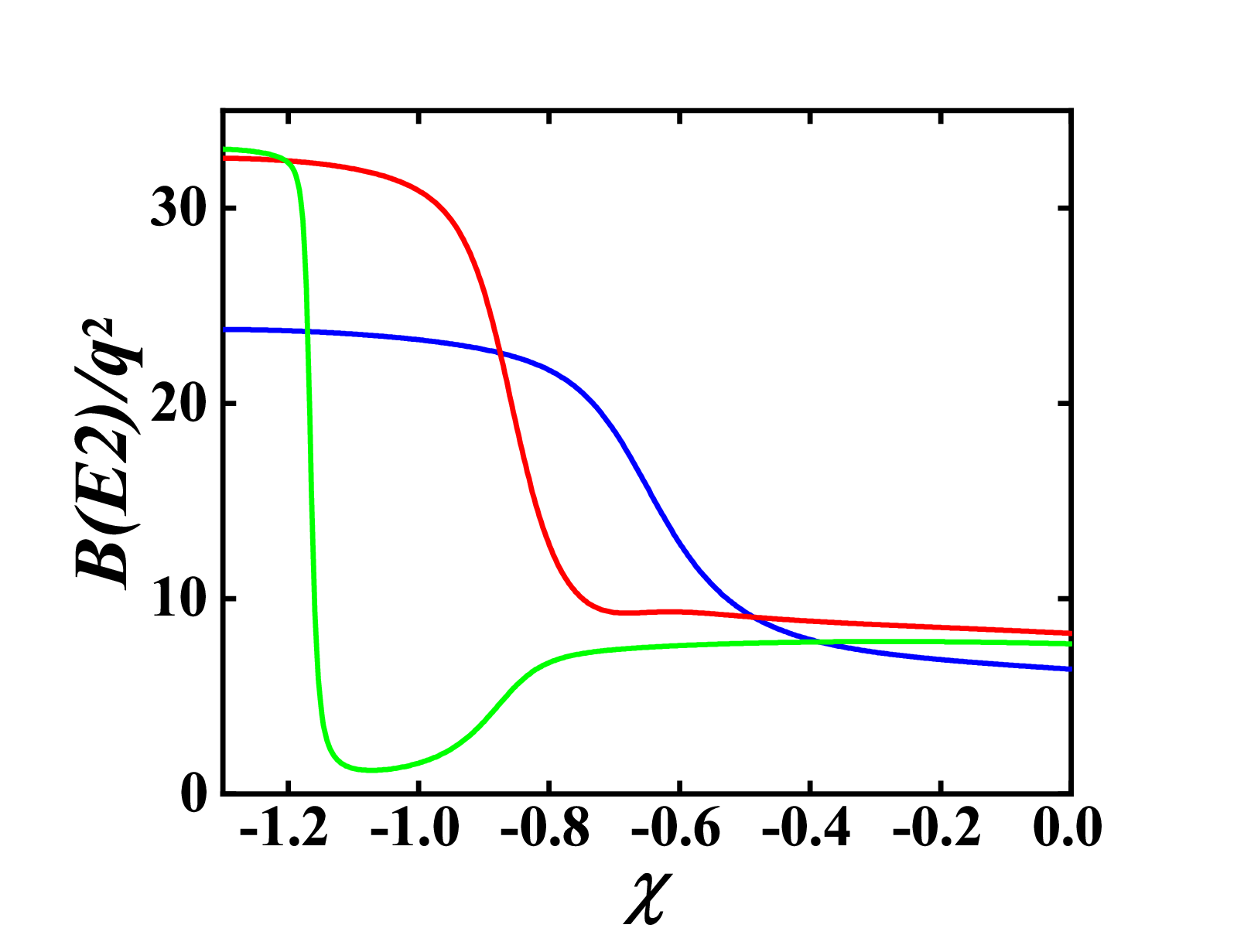}
\caption{(a) The evolutional behaviors of the partial low-lying levels as a function of $\chi$; (b) The evolutional behaviors of the $B(E2; 2_{1}^{+}\rightarrow 0_{1}^{+})$ (blue line) , $B(E2; 4_{1}^{+}\rightarrow 2_{1}^{+})$ (red line), $B(E2; 6_{1}^{+}\rightarrow 4_{1}^{+})$ (green line) as a function of $\chi$. The parameters are deduced from \cite{Wang20}.}
\end{figure}

\textbf{Level-crossing mechanism was first proposed in \cite{Wang20}, which is also} the first theoretical explanation for the B(E2) anomaly in realistic nuclei. In this explanation, the SU(3) third-order interaction $[\hat{L} \times \hat{Q} \times \hat{L}]$ plays a key role, where $\hat{Q}$ is the SU(3) quadrupole operator. In the SU(3) symmetry limit, the $[\hat{L} \times \hat{Q} \times \hat{L}]$ interaction can lower the energy of a $4^{+}$ state in the SU(3) irreducible representation (irrep) $(2N-8,4)$ and increase the energy of the $4^{+}$ state in the SU(3) irrep $(2N,0)$. Therefore, \textbf{the two $4^{+}$ states can crossover with each other and} the former \textbf{level} can be lower than the latter \textbf{one}, and the ratio $B_{4/2}$ is zero. This situation occurs within the SU(3) symmetry limit and is a level-crossing phenomenon.

\textbf{In this paper, we mainly discuss $^{166}$Os, and its boson number is $N=7$. Thus we introduce these fundamental concepts in the level-crossing mechanism or the general explanatory framework with $N=7$, and using the parameters in \cite{Wang20}. The cases with $N=8$, 9 have been discussed in \cite{Wang20,Cheng25,Li25}.
}

\textbf{Whether this B(E2) mechanism is related to the SU(3) symmetry is an important problem. If the B(E2) anomaly in an extended IBM model is also anomalous in its SU(3) symmetry limit, the explanation is regarded to be related to the SU(3) symmetry. In \cite{Wang20}, this point was noticed, but not emphasized. SU(3) analysis is a useful technique to study the relationship between one B(E2) mechanism and the SU(3) symmetry. For one Hamiltonian used to explain the B(E2) anomaly, it can be divided into two parts: one is related to the SU(3) symmetry limit, and one other is not related. For the SU(3) symmetry limit part, let the parameter $\eta$ in front of the third-order interaction $[\hat{L} \times \hat{Q} \times \hat{L}]^{(0)}$ change gradually, and observe whether the $4_{1}^{+}$ state can intersect with one other higher $4^{+}$ state and other level-crossing phenomena.}

\textbf{Fig. 1(a) shows the evolutional behaviors when the parameter $\eta$ decreases from 0 for the low-lying $0^{+}$, $2^{+}$, $4^{+}$ and $6^{+}$ states. We can observe that, the first $4^{+}$ state crossovers with one other higher $4^{+}$ state at $\eta=-15.20$ keV (real red lines and black circle). Before this crossover point, the $6_{1}^{+}$ state first intersects with one other higher $6^{+}$ state at $\eta=-10.04$ keV (real green lines and black circle). When $\eta$ further decreases, the first $2^{+}$ state can also crossover with one other higher $2^{+}$ state at $\eta=-30.72$ keV (real blue lines and black circle). In the SU(3) symmetry limit, if two levels belong to two different SU(3) irreps $(\lambda, \mu)$, the B(E2) transitions between the two levels must be zero. After the crossover point of the two $4^{+}$ states and before the crossover point of the two $2^{+}$ states (from $-$15.20 keV to $-$30.72 keV), B(E2) anomaly exists for the $B_{4/2}$ value is 0.}

\textbf{The $B_{4/2}$ values in the U(5) symmetry limit and the O(6) symmetry limit are both normal \cite{Wangtao}, that $B_{4/2}>1.0$. Thus if $B_{4/2}=0$ in the SU(3) symmetry limit, the realistic $B_{4/2}<1.0$ value can be obtained when moving towards the U(5) symmetry limit and the O(6) symmetry limit. However the two recovery mechanisms are very different. In Fig. 1, the crossover of the $4_{1}^{+}$ state and one other higher $4^{+}$ state induces the B(E2) anomaly, thus when moving towards the U(5) symmetry limit, the two $4^{+}$ states have level-anticrosssing phenomenon, unwinding the crossover in the SU(3) symmetry limit.}

\textbf{However when moving towards the O(6) symmetry limit, the unwinding phenomenon can not occur \cite{Li25}. } \textbf{Fig. 2 shows the evolutional behaviors from the SU(3) symmetry limit to the O(6) symmetry limit when $\eta=-20.0$ keV (here the parameter changes from $-\frac{\sqrt{7}}{2}$ to $-0.7$). Clearly the evolutional behaviors of the two $4_{1}^{+}$ and $4_{2}^{+}$ states have not level-anticrossing. Instead the B(E2) anomaly results from the level-anticrossing of the two $2_{1}^{+}$ and $2_{2}^{+}$ states (real blue lines and black circle).}

\begin{figure}[tbh]
\includegraphics[scale=0.33]{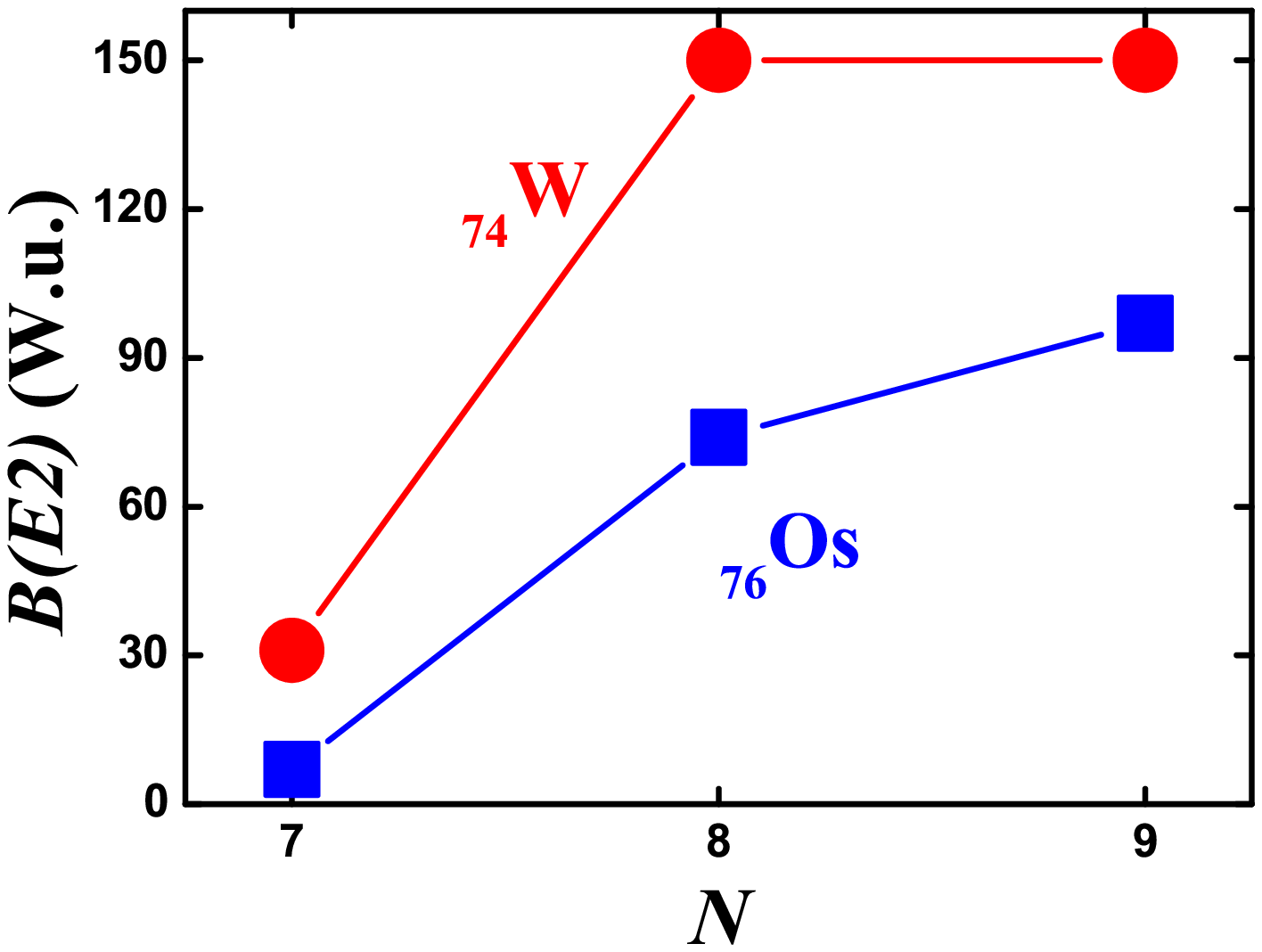}
\caption{Experimental $B(E2;2_{1}^{+}\rightarrow0_{1}^{+})$ values in $^{162-166}$W and $^{166-170}$Os as a function of the boson number $N$. These values are from \cite{Garahn16,Joss17,Goasduff19,Stolze21,Doncel17}.}
\end{figure}

\begin{table}[tbh]
\caption{\label{table:expee} $B(E2;2^{+}_{1}\rightarrow0^{+}_{1})$ values of 22 nuclei with $N=7$. The unit is W.u.. These values are from \cite{Stolze21,Doncel17,ensdf,Xe114,Moller05,Zhu17,Daniel17,Bello18,Esmayl18,Kisyov22}.}
\setlength{\tabcolsep}{2.5mm}{
\begin{tabular}{ccccccc}
\hline
\hline
  nucleus           &$B(E2;2^{+}_{1}\rightarrow0^{+}_{1})$         \ &nucleus              \ &$B(E2;2^{+}_{1}\rightarrow0^{+}_{1})$                \\
 \hline
$^{146}$Gd          &            $>$0.59                    \ & $^{118}$Sn         \ & 12.1(5)           \\
$^{114}$Sn         &            15(3)                   \ &         \ &           \\
 \hline
$^{194}$Hg         &           $39^{+9}_{-6}$                   \ &  $^{122}$Te      \ &    36.92(25)          \\
$^{118}$Cd         &        33(3)                      \  &  $^{114}$Te               \ & 34.0(30)  \\
   $^{110}$Cd       &             27.0(8)                 \ & \ &   \\
 \hline
$^{194}$Pt  &   49.5(20)     \ &  $^{146}$Nd     \ &    31.9(4)    \\
$^{138}$Nd   &     36(1)      \ &  $^{126}$Xe      \ &  56(5)      \\
$^{114}$Xe   &    62(4)      \ &   $^{106}$Pd     \ &    44.3(15)      \\
 \hline
 $^{194}$Os         &        45(16)              \ &   $^{166}$Os      \ &  \textbf{7(4)}    \\
$^{162}$W     &  \textbf{31(13)}            \ &  $^{146}$Ce     \ &   43(5)      \\
$^{146}$Ba     &  59.7(19)     \ & $^{134}$Ce    \ & 50.8(41) \\
$^{130}$Ba   &     57.9(17)    \ & $^{102}$Ru \ &  44.6 (7)            \\
\hline
\hline
\end{tabular}}
\end{table}

In \cite{Pan24}, it was found that, even \textbf{if} the $B_{4/2}$ value is larger than 1.0 in the SU(3) symmetry limit (here $\eta>15.20$ keV), when moving towards the O(6) symmetry limit, the B(E2) anomaly can also occur. \textbf{Fig. 3 shows the evolutional behaviors from the SU(3) symmetry limit to the O(6) symmetry limit when $\eta=-10.0$ keV. Clearly in the SU(3) symmetry limit, the $B_{4/2}$ value is normal. When $\chi$ increases, the $6_{1}^{+}$ and $6_{2}^{+}$ states first have level-anticrossing (real green lines and black circle), and then $4_{1}^{+}$ and $4_{2}^{+}$ states (real red lines and black circle), and last the $2_{1}^{+}$ and $2_{2}^{+}$ states (real blue lines and black cirlce). Thus the B(E2) anomaly can occur. This new mechanism has been combined into previous level-crossing mechanism and a general explanatory framework has been obtained. Detailed discussion can be seen in \cite{Li25}.
}

\textbf{The three different mechanisms discussed here will be used to fit the $B(E2;2^{+}_{1}\rightarrow0^{+}_{1})$ anomaly in $^{166}$Os and the B(E2) anomalies in $^{168,170}$Os.
}

\section{$B(E2;2^{+}_{1}\rightarrow0^{+}_{1})$ anomaly in $^{166}$Os}

In Fig. 4, the evolutional behavior of the $B(E2;2_{1}^{+}\rightarrow0_{1}^{+}$) values in $^{166-170}$Os is shown. When the boson number decreases from 9 to 7, the value decreases from 97(9) W.u., normally to 74(13) W.u., and then, in a sudden way, to 7(4) W.u. The $B(E2;2^{+}_{1}\rightarrow0^{+}_{1})$ value in $^{166}$Os is almost 10 times smaller than the one in $^{168}$Os while the boson number is only one less. Such a very small $B(E2;2^{+}_{1}\rightarrow0^{+}_{1})$ value, in general, can only occur in magic nuclei. When moving away from the magic nuclei, this value increases greatly. If $N\geq5$, the nucleus can have a deformed shape, and the $B(E2;2^{+}_{1}\rightarrow0^{+}_{1})$ value is large. The two adjacent nuclei $^{168, 170}$Os are really so. Similar evolutional behavior can be also observed in $^{162-166}$W. The $B(E2;2^{+}_{1}\rightarrow0^{+}_{1})$ value in $^{162}$W is 31(13) W.u. \cite{Doncel17}, which is almost 5 times smaller than the value 150(100) W.u. in $^{164}$W \cite{Doncel17}.

In Table I, the $B(E2;2_{1}^{+}\rightarrow0_{1}^{+}$) values of 22 nuclei with $N=7$ are shown. From the top, the three nuclei $^{146}$Gd, $^{118}$Sn and $^{114}$Sn in the first group are magic nuclei (proton or neutron boson number is 0). The $B(E2;2_{1}^{+}\rightarrow0_{1}^{+}$) values are small, but the values in $^{114,118}$Sn are still larger than the one in $^{166}$Os. The proton or neutron boson number of the five nuclei in the second group is 1, and the average value of the five nuclei is 34.0 W.u., much larger than the ones of the magic nuclei in the first group. Next, the proton or neutron boson number of the six nuclei in the third group is 2, and the average value is 46.6 W.u., larger than the one of the nuclei in the second group.

Last, the proton or neutron boson number of the eight nuclei in the fourth group is 3, and the average value except for $^{166}$Os and $^{162}$W is 50.2 W.u., larger than the one of the nuclei in the third group. This is the normal trend. If \textbf{including} $^{166}$Os and $^{162}$W, the average value is 38.6 W.u., smaller than the one 46.6 W.u. in the third group. The \textbf{normal} average value 50.2 W.u. can be also deduced from Fig. 1 from normal extrapolation. The $B(E2;2_{1}^{+}\rightarrow0_{1}^{+}$) value in $^{166}$Os is very small, almost 7 times smaller than this average value.

\textbf{Through above discussions,} the $B(E2;2_{1}^{+}\rightarrow0_{1}^{+}$) anomaly in $^{166}$Os really exists.

\section{Hamiltonian}

\begin{figure}[tbh]
\includegraphics[scale=0.33]{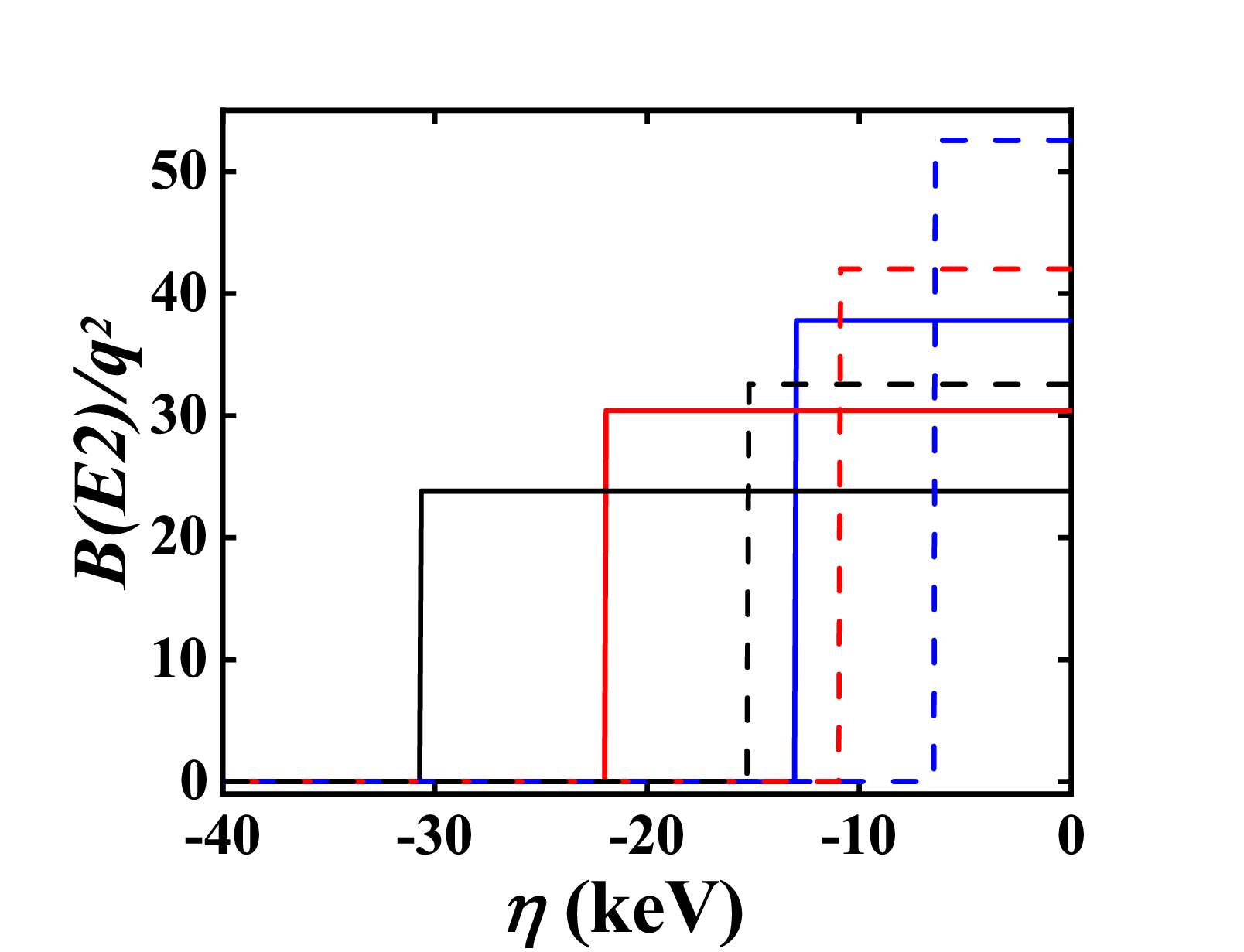}
\caption{The evolutional behaviors of the $B(E2;2_{1}^{+}\rightarrow0_{1}^{+})$ (solid lines), $B(E2;4_{1}^{+}\rightarrow2_{1}^{+})$ (dashed lines) values as a function of $\eta$ for $N=9$ (blue lines), $N=8$ (red lines) and $N=7$ (black lines). The parameters are deduced from \cite{Wang20}.}
\end{figure}

In \cite{Li25}, a general explanatory framework for the B(E2) anomaly was proposed based on the SU(3) analysis up to the SU(3) third-order interactions. This Hamiltonian is as follows
\begin{eqnarray}
\hat{H}&=&\varepsilon_{d}\hat{n}_{d}-\kappa\hat{Q}_{\chi} \cdot \hat{Q}_{\chi}+\zeta[\hat{Q}_{\chi} \times \hat{Q}_{\chi} \times \hat{Q}_{\chi}]^{(0)}      \nonumber\\
&&            +\eta[\hat{L} \times \hat{Q}_{\chi} \times \hat{L}]^{(0)}+f\hat{L}^{2},
\end{eqnarray}
here $\varepsilon_{d}$, $\kappa$, $\zeta$, $\eta$  and $f$ are five fitting parameters. $\hat{n}_{d}=d^{\dag} \cdot \tilde{d}$ is the $d$ boson number operator, and $\hat{Q}_{\chi}= (d^{\dag}s + s^{\dag}\tilde{d}) + \chi(d^{\dag} \times \tilde{d})$ is the general quadrupole operator ($-\frac{\sqrt{7}}{2}\leq \chi \leq 0$). $\hat{Q}=\hat{Q}_{-\frac{\sqrt{7}}{2}}$. If $\varepsilon_{d}=0$ and $\chi=-\frac{\sqrt{7}}{2}$, this Hamiltonian is the one for the SU(3) analysis. The $-\hat{Q}\cdot \hat{Q}$ interaction can describe the prolate shape, and the $-[\hat{Q} \times \hat{Q} \times \hat{Q}]^{(0)}$ interaction can describe the oblate shape \cite{Fortunato11}. The third-order interaction $[\hat{L} \times \hat{Q} \times \hat{L}]^{(0)}$ is vital for the emergence of the B(E2) anomaly.

For understanding the B(E2) anomaly, the B(E2) values are necessary. The $E2$ operator is defined as
\begin{eqnarray}
\hat{T}(E2)=q\hat{Q}_{\chi},
\end{eqnarray}
where $q$ is the boson effective charge. The evolutional behaviors of the $B(E2;4_{1}^{+}\rightarrow2_{1}^{+})$ and $B(E2;2_{1}^{+}\rightarrow0_{1}^{+})$ values are discussed. Here $q=Nq_{0}$ is used. \textbf{When discussing the higher-order interactions, the simple form in (2) is usually used \cite{Isacker99,Arias00,zhang14}. If more accurate results are obtained, higher-order interaction $[\hat{Q}_{\chi} \times \hat{Q}_{\chi}]^{(2)}$ should be considered \cite{Iachello87,Heyde84}. In the existing discussions with the SU3-IBM, we found that the simple form in (2) is sufficiently enough \cite{WangPt,WangPd106}.}

\textbf{One may doubt that whether the boson number $N$ used here is applicable. In a recent paper on the boson number odd-even effect in $^{196-204}$Hg \cite{WangHg}, it has proved that the boson number $N$ must be the valence nucleon-pair number, which validates the boson number assumption in the IBM.}

\section{SU(3) analysis}

\begin{table}[tbh]
\caption{\label{table:expee} The fitting parameters of the four results for $^{166,168,170}$Os. The unit is keV except for $\chi$.}
\setlength{\tabcolsep}{1.0mm}{
\begin{tabular}{ccccccc}
\hline
\hline
  Res. 1      & $\chi $ \ &$\varepsilon_{d} $  \ & $\kappa$\  &  $\zeta$ \  &$\eta$\  & $f$ \\
 \hline
 $^{170}$Os &    0     \ & 306     \ & 30.09  \ & 3.79  \ & $-10.38$  \  &18.66\\
 $^{168}$Os &    0     \ & 345.72  \ & 31.72  \ &3.99   \  &$-14.69$  \  &19.67\\
 $^{166}$Os &    0     \ & 31.2    \ &10.76  \ &1.36    \  &$-11.91$  \  &50.20\\
 \hline
 \hline
 Res. 2      & $\chi $ \ &$\varepsilon_{d} $  \ & $\kappa$\  &  $\zeta$ \  &$\eta$\  & $f$ \\
 \hline
 $^{170}$Os &  $-1.1192$     \ & 0  \ & 21.60  \ & $-5.01$  \ &$-8.91$   \ &27.79 \\
 $^{168}$Os &  $-1.0266$     \ & 0  \ & 30.27  \ &$-5.01$   \ &$-12.01$  \ &33.64 \\
 $^{166}$Os &  $-1.3044$     \ & 0  \ & 21.60  \ & $-3.59$  \ &$-13.88$  \ &41.57\\
 \hline
 \hline
 Res. 3      & $\chi $ \ &$\varepsilon_{d} $  \ & $\kappa$\  &  $\zeta$ \  &$\eta$\  & $f$ \\
 \hline
 $^{170}$Os &  $-1.0581$    \ & 0   \ & 79.31  \ & $-13.13$\  & $-16.12$  \  &5.06\\
 $^{168}$Os &  $-0.9551$    \ & 0   \ & 85.64  \ &$-14.18$ \  & $-21.63$  \  &2.77\\
 $^{166}$Os &  $-1.0504$    \ & 0   \ & 29.26  \ &$-4.84$  \  & $-18.45$  \  &45.76\\
 \hline
 \hline
 Res. 4      & $\chi $ \ &$\varepsilon_{d} $  \ & $\kappa$\  &  $\zeta$ \  &$\eta$\  & $f$ \\
 \hline
 $^{170}$Os &  $-1.1562$    \ & 0   \ & 120.69  \ & $-19.99$  \  & $-24.54$  \  & $-21.49$   \\
 $^{168}$Os &  $-1.1324$    \ & 0   \ & 97.67   \ & $-16.17$  \  & $-26.71$  \  & $-11.73$   \\
 $^{166}$Os &  $-1.1192$    \ & 0   \ & 25.52   \ & $-4.23$   \  & $-15.70$  \  & 42.84    \\
\hline
\hline
\end{tabular}}
\end{table}

\textbf{Following the ideas in section II,} we perform the SU(3) analysis \textbf{for $N=7, 8, 9$}. In Fig. 5, the evolutional behaviors of the $B(E2;2_{1}^{+}\rightarrow 0_{1}^{+}$) values (real lines), $B(E2;4_{1}^{+}\rightarrow2_{1}^{+})$ values (dashed lines) as a function of $\eta$ are presented. Other parameters are $\varepsilon_{d}=0$ keV, $\chi=-\frac{\sqrt{7}}{2}$, $\kappa=30.09$ keV, $\zeta=3.79$ keV, $f=18.66$ keV \cite{Wang08}. The boson numbers $N$ are 7 for $^{166}$Os (black lines), 8 for $^{168}$Os(red lines) and 9 for $^{170}$Os (blue lines) respectively. The SU(3) irrep of the ground state is $(2N,0)$, which is the prolate shape. Thus the B(E2) values are the largest among all the SU(3) irreps $(\lambda,\mu)$.

For different $N$, the parameters $\eta$ for the emergences of $B(E2;4_{1}^{+}\rightarrow 2_{1}^{+})=0$, $B(E2;2_{1}^{+}\rightarrow 0_{1}^{+})=0$ are different, and if $N$ decreases, the parameters decrease too.

The validity of the parameter setting of the effective charge $q=Nq_{0}$ needs to be explained here. Under normal circumstances, the $B(E2;2_{1}^{+}\rightarrow 0_{1}^{+})$ value in $^{166}$Os should be around 50.2 W.u. (the average value discussed in section \textbf{III}). The $B(E2;2_{1}^{+}\rightarrow 0_{1}^{+})$ values in $^{168,170}$Os are 74(13) W.u. and 97(9) W.u. respectively. \textbf{Thus the normal ratio for $^{166-170}$Os is 50.2:74:97.} In Fig. 5, in the SU(3) analysis, the ratio of the $B(E2;2_{1}^{+}\rightarrow 0_{1}^{+})$ values for $N=7,8,9$ is 23.8:30.4:37.8 \textbf{or 61:78:97.} \textbf{If $q$ is the same for $^{166-170}$Os, the experimental data can not be obtained.} If $q=Nq_{0}$, the ratio of the $B(E2;2_{1}^{+}\rightarrow 0_{1}^{+})$ values of the fit for $^{166-170}$Os is $61\times7^{2}:78\times8^{2}:97\times9^{2}$=36.9:61.6:97. When the $\hat{n}_{d}$ interaction is added or the parameter $\chi$ changes from $-\frac{\sqrt{7}}{2}$ to 0, \textbf{the normal ratio 50.2:74:97 can be obtained}. Thus the setting $q=Nq_{0}$ is reasonable, \textbf{and the very small $B(E2;2_{1}^{+}\rightarrow 0_{1}^{+})$ value in $^{166}$Os results from level-crossing. }

\section{Results}

Now, we fit $^{166,168,170}$Os based on the results in Fig. 5. In this paper, we present four results for $^{166-170}$Os from \textbf{three} different mechanisms \textbf{shown in section II}. Table II presents the fitting parameters of the four results.

\begin{table}[tbh]
\caption{\label{table:expee} Experimental energy values and fitted data of the four results for some states along the yrast band for $^{166-170}$Os. The unit is keV. }
\setlength{\tabcolsep}{2.4mm}{
\begin{tabular}{ccccccc}
\hline
\hline
 $^{170}$Os   &Exp.    \ &Res. 1     \ &Res. 2     \ &Res. 3      \ & Res. 4   \\
 \hline
$E_{2_{1}^{+}}$& 287    \ & 284       \ & 287       \ & 287        \ & 287      \\
$E_{4_{1}^{+}}$& 750    \ & 733       \ & 750       \ & 750        \ & 901      \\
$E_{6_{1}^{+}}$& 1325   \ & 1237      \ & 1476      \ & 1422       \ & 1245     \\
$E_{8_{1}^{+}}$& 1946   \ & 1960      \ & 2558      \ & 2413       \ & 2621     \\
 \hline
 \hline
 $^{168}$Os &Exp.      \ &Res. 1      \ &Res. 2     \ &Res. 3   \ & Res. 4  \\
 \hline
$E_{2_{1}^{+}}$& 341    \ & 343       \ & 343       \ & 341        \ & 342       \\
$E_{4_{1}^{+}}$& 857    \ & 973       \ & 867       \ & 849        \ & 1085      \\
$E_{6_{1}^{+}}$& 1499   \ & 1549      \ & 1697      \ & 1511       \ & 1500      \\
$E_{8_{1}^{+}}$& 2223   \ & 2431      \ & 2987      \ & 2537       \ & 2225      \\
\hline
\hline
 $^{166}$Os &Exp.       \ &Res. 1      \ &Res. 2     \ &Res. 3   \ & Res. 4  \\
 \hline
$E_{2_{1}^{+}}$& 432    \ & 432       \ & 411       \ & 440        \ & 429      \\
$E_{4_{1}^{+}}$& 1020   \ & 1020      \ & 1010      \ & 1046       \ & 1019     \\
$E_{6_{1}^{+}}$& 1724   \ & 1975      \ & 1948      \ & 2074       \ & 1990     \\
$E_{8_{1}^{+}}$& 2350   \ & 3653      \ & 3576      \ & 3873       \ & 3663     \\
\hline
\hline
\end{tabular}}
\end{table}

The idea of the fit for any result is as follows. We first fit the B(E2) anomaly in $^{170}$Os, for which a lot of experiences have been accumulated. Based on the results (blue lines) in Fig. 5, the parameter $\eta$ is determined, then the $\hat{n}_{d}$ interaction is added or the parameter $\chi$ changes from $-\frac{\sqrt{7}}{2}$ to 0. The $B_{4/2}$ value matching the experimental one in $^{170}$Os can be found. Let the energy of the $2_{1}^{+}$ state be equal to the experimental one, all the parameters should be multiplied by the same number, so that the parameters can be determined, see Table II. Let the $B_{4/2}$ value in $^{170}$Os be equal to the experimental one, the parameter $q_{0}$ can be determined, and then the effect charge $q$ is obtained. For each result, the parameter $q_{0}$ is the same. This makes the change of the effective charge $q$ not too large for $^{168-170}$Os and in line with the actual situation. Using similar method, the parameters of $^{168}$Os and $^{166}$Os can be obtained successively.

\subsection{Result 1}

For result 1, the $\hat{n}_{d}$ interaction is added (\textbf{the first mechanism}). $^{168,170}$Os has B(E2) anomaly. The parameter $\eta$ is chosen from the region satisfying $B(E2;2^{+}_{1}\rightarrow0^{+}_{1})\neq0$ and $B(E2;4^{+}_{1}\rightarrow2^{+}_{1})=0$. For $^{166}$Os, the $B(E2;2^{+}_{1}\rightarrow0^{+}_{1})$ value is very small, so the parameter $\eta$ is determined from the region satisfying $B(E2;2^{+}_{1}\rightarrow0^{+}_{1})=0$ and $B(E2;4^{+}_{1}\rightarrow2^{+}_{1})=0$. This possibility is unique if the results are obtained by adding the $\hat{n}_{d}$ interaction.

\begin{table}[tbh]
\caption{\label{table:expee}  Experimental B(E2) values and fitted data of the four results for the $B(E2;2^{+}_{1}\rightarrow0^{+}_{1})$, $B(E2;4^{+}_{1}\rightarrow2^{+}_{1})$, $B(E2;6^{+}_{1}\rightarrow4^{+}_{1})$, $B(E2;8^{+}_{1}\rightarrow6^{+}_{1})$ values for $^{166-170}$Os. The unit is W.u.. For result $1-4$ of $^{170}$Os, $q_{0}$ are 0.0924 eb, 0.1495 eb, 0.0942 eb and 0.0755 eb, respectively.}
\setlength{\tabcolsep}{0.8mm}{
\begin{tabular}{ccccccccc}
\hline
\hline
$^{170}$Os                                  &Exp.      \ &Res. 1      \ &Res. 2     \ &Res. 3   \ & Res. 4  \\
 \hline
 $B(E2;2^{+}_{1}\rightarrow0^{+}_{1})$  \   & 97(9)       \ & 96         \ & 97        \ &97       \ & 97      \\
 $B(E2;4^{+}_{1}\rightarrow2^{+}_{1})$  \   & 38$^{+18}_{-7}$       \ & 38         \ & 38        \ &38       \ & 38      \\
 $B(E2;6^{+}_{1}\rightarrow4^{+}_{1})$  \   &          \ & 34         \ & 42        \ &41       \ & 14      \\
 $B(E2;8^{+}_{1}\rightarrow6^{+}_{1})$  \   &          \ & 22         \ & 34        \ &34       \ & 15      \\
$\emph{ B}_{4/2}$                           \   & 0.39    \ & 0.39      \ & 0.39     \ &0.40    \ & 0.39    \\
 \hline
\hline
$^{168}$Os                                  &Exp.       \ &Res. 1      \ &Res. 2     \ &Res. 3   \ & Res. 4  \\
 \hline
 $B(E2;2^{+}_{1}\rightarrow0^{+}_{1})$  \   & 74(13)       \ & 75         \ & 76        \ &74       \ & 63      \\
 $B(E2;4^{+}_{1}\rightarrow2^{+}_{1})$  \   & 25(13)       \ & 25         \ & 25        \ &25       \ & 22      \\
 $B(E2;6^{+}_{1}\rightarrow4^{+}_{1})$  \   &          \ & 18         \ & 28        \ &26       \ & 6       \\
 $B(E2;8^{+}_{1}\rightarrow6^{+}_{1})$  \   &          \ & 11         \ & 22        \ &21       \ & 6       \\
$\emph{ B}_{4/2}$                           \   & 0.338    \ & 0.340      \ & 0.333     \ &0.340    \ & 0.346   \\
 \hline
\hline
 $^{166}$Os                                 &Exp.     \ &Res. 1      \ &Res. 2     \ &Res. 3   \ & Res. 4  \\
 \hline
 $B(E2;2^{+}_{1}\rightarrow0^{+}_{1})$  \   & 7(2)        \ & 7.2        \ & 7.2       \ &7.0      \ & 7.1     \\
 $B(E2;4^{+}_{1}\rightarrow2^{+}_{1})$  \   &          \ & 0.3        \ & 0.1       \ &3.3      \ & 1.1     \\
 $B(E2;6^{+}_{1}\rightarrow4^{+}_{1})$  \   &          \ & 0.4        \ & 0.1       \ &3.7      \ & 1.6     \\
 $B(E2;8^{+}_{1}\rightarrow6^{+}_{1})$  \   & 1.4(5)      \ & 1.0        \ & 2.1       \ &2.8      \ & 1.4     \\
 $\emph{B}_{8/2}$                           \   & 0.20    \ & 0.14      \ & 0.29     \ &0.40    \ & 0.20    \\
\hline
\hline
\end{tabular}}
\end{table}

It should be noticed that, the choice of the parameters is nearly robust. The fitting values of the result 1 for $^{166-170}$Os can be seen in Table \textbf{III} (the energies of the $2_{1}^{+}$, $4_{1}^{+}$, $6_{1}^{+}$ and $8_{1}^{+}$ states) and Table \textbf{IV} (the values of the E2 transitions $B(E2;2_{1}^{+}\rightarrow0_{1}^{+})$, $B(E2;4_{1}^{+}\rightarrow2_{1}^{+})$, $B(E2;6_{1}^{+}\rightarrow4_{1}^{+})$, $B(E2;8_{1}^{+}\rightarrow6_{1}^{+})$). The B(E2) values agree with the experimental ones. For $^{166}$Os, not only the $B(E2;2_{1}^{+}\rightarrow0_{1}^{+})$ value but also the $B(E2;8_{1}^{+}\rightarrow6_{1}^{+})$) value has been measured. The experimental $B(E2;8_{1}^{+}\rightarrow6_{1}^{+})$) value is 1.4(5) W.u., and the theoretical value is 1.0 W.u. Thus the fitting effect of the result 1 is very good.

Ref. \cite{Wang20} proposed the SU3-IBM and described the B(E2) anomaly in $^{170}$Os, which opened a new understanding of the nuclear structure. Now we show that this mechanism can still describe the B(E2) anomaly in $^{168}$Os and the $B(E2;8_{1}^{+}\rightarrow6_{1}^{+})$) anomaly in $^{166}$Os. Thus the level-crossing explanation is more successful.

\subsection{Result 2}

For results $2-4$, we set $\varepsilon_{d}$ = 0 and slightly adjust $\chi$, which was proposed by \cite{Pan24}. For result 2, we also select the parameters $\eta$ for fitting $^{166-170}$Os in the anomalous region of Fig. 5 (\textbf{the second mechanism in Fig. 2}). From Table \textbf{III} and Table \textbf{IV}, the fitting effect of the result 2 is also very good except that the $B(E2;8_{1}^{+}\rightarrow6_{1}^{+})$) value in $^{166}$Os is somewhat larger.

\subsection{Result 3 and 4}

When the parameter $\chi$ changes from $-\frac{\sqrt{7}}{2}$ to 0, Ref. \cite{Pan24} found new possibilities. If the $B_{4/2}$ value is larger than 1.0 in the SU(3) analysis, when $\chi$ changes, the B(E2) anomaly can be also obtained (\textbf{the third mechanism in Fig. 3}). These has been discussed in detail in \cite{Pan24,Li25}. Level-anticrossing of the $4_{1}^{+}$ state and one other $4^{+}$ state can occur. Thus different from result 1 and 2, in this case, there are two parameter points satisfying the experimental one in $^{170}$Os, thus we have result 3 and 4.

Result 3 chooses the $B_{4/2}$ value towards the O(6) side in $^{170}$Os. From Table \textbf{III} and Table \textbf{IV}, the fitting effect is still very good except that the $B(E2;8_{1}^{+}\rightarrow6_{1}^{+})$) value in $^{166}$Os is more larger than the ones in result 1 and 2. The $B(E2;4_{1}^{+}\rightarrow2_{1}^{+})$), $B(E2;6_{1}^{+}\rightarrow4_{1}^{+})$) values in $^{166}$Os are much larger than the ones in result 1 and 2.

Result 4 chooses the $B_{4/2}$ value towards the SU(3) side in $^{170}$Os. This possibility was pointed out in \cite{Li25}. From Table \textbf{III} and Table \textbf{IV}, the fitting effect is still very good. The $B(E2;4_{1}^{+}\rightarrow2_{1}^{+})$) values in $^{168,170}$Os are much smaller than the ones in result 1 and 2.

\subsection{Brief discussions}

\textbf{For Hamiltonian (1), only the third-order interactions are considered, and the fourth-order interactions are not added. The reason for this is that the functions of these fourth-order interactions on the B(E2) anomaly are not discussed clearly because the effects may be very complicated, and detailed studies are needed in future. In \cite{Wang25}, it was found that the fourth-order interactions can affect the positions of the levels greatly, but the B(E2) values slightly. When discussing the Cd nuclei, the fourth-order interactions can make the energy levels fit well. In a recent paper on $^{106}$Pd \cite{WangPd106}, this result can be also obtained. Thus in following studies, the introduction of the fourth-order interactions may make the energy levels in Table III fit better, especially the $8_{1}^{+}$ state.}

\textbf{It should be stressed that the small $B(E2;8_{1}^{+}\rightarrow6_{1}^{+})$ value is not an accident result. In the level-crossing explanation, when $[\hat{L} \times \hat{Q} \times \hat{L}]^{(0)}$ decreases, the $B(E2;8_{1}^{+}\rightarrow6_{1}^{+})$ anomaly first occurs, and then $B(E2;6_{1}^{+}\rightarrow4_{1}^{+})$ anomaly, $B(E2;4_{1}^{+}\rightarrow2_{1}^{+})$ anomaly and $B(E2;2_{1}^{+}\rightarrow0_{1}^{+})$ anomaly. Thus the small $B(E2;2_{1}^{+}\rightarrow0_{1}^{+})$) value implies the emergence of the small $B(E2;8_{1}^{+}\rightarrow6_{1}^{+})$) value, and our fits prove this result.}

\textbf{In this paper, we show that the level-crossing mechanism or the general explanatory framework really explain the $B(E2;2_{1}^{+}\rightarrow0_{1}^{+})$ anomaly in $^{166}$Os. However the best fitting conditions are not discussed, because the fourth-order interaction effects are not clear. In Table II, the fitting parameters for $^{166-170}$Os are not so natural, and change abruptly, so a better fitting effect is needed. This will be discussed in next paper. }

\section{Conclusion}

This paper employs the SU(3) analysis technique to comparatively analyze the three mechanisms that cause the B(E2) anomaly to explain the very small $B(E2;2_{1}^{+}\rightarrow0_{1}^{+})$ anomaly in $^{166}$Os. Appropriate parameters are selected to fit the $^{166,168,170}$Os. The fitting results closely approximate the experimental data. Not only the B(E2) anomaly in $^{168,170}$Os but also the $B(E2;2_{1}^{+}\rightarrow0_{1}^{+})$ anomaly in $^{166}$Os are described. This implies that the level-crossing or level-anticrossing mechanism is more convincing. There are many possible mechanisms for describing the B(E2) anomaly, here we show that the general explanatory framework in \cite{Li25} is really useful.

\end{document}